\documentclass[aps,prc,showpacs,twocolumn,groupeddaddress]{revtex4} 	%%% revtex4
\newcommand{\bfi}[1]{\mbox{\boldmath $#1$}}
\newcommand{\Lower}[1]{\smash{\lower 1.5ex \hbox{#1}}}
\newcommand{\LN}{$\Lambda N$}
\newcommand{\LL}{$\Lambda\Lambda$}
\newcommand{\pnLL}{$pn\Lambda\Lambda$}

\newcommand{\LNSN}{$\Lambda N-\Sigma N$}

\newcommand{\BL}{$B_\Lambda$}
\newcommand{\BLL}{$B_{\Lambda\Lambda}$}

\newcommand{\HII}{$^2$H}
\newcommand{\HIII}{$^3$H}
\newcommand{\HeIII}{$^3$He}
\newcommand{\HeIV}{$^4$He}
\newcommand{\HIIIL}{$_\Lambda^3$H}
\newcommand{\HIIILs}{$_\Lambda^3$H$^\ast$}
\newcommand{\HIVL}{$_\Lambda^4$H}

\newcommand{\HIVLs}{$_\Lambda^4$H$^\ast$}

\newcommand{\HeVL}{$_\Lambda^5$He}
\newcommand{\HIVLL}{$_{\Lambda\Lambda}^{\ \ 4}$H}

\newcommand{\HeVILL}{$_{\Lambda\Lambda}^{\ \ 6}$He}

\makeatletter
\def\gsim{\compoundrel>\over\sim}

\def\compoundrel#1\over#2{\mathpalette\compoundreL{{#1}\over{#2}}}
\def\compoundreL#1#2{\compoundREL#1#2}
\def\compoundREL#1#2\over#3{\mathrel
	{\vcenter{\hbox{$\m@th\buildrel{#1#2}\over{#1#3}$}}}}
\makeatother

\usepackage{graphicx}%%% Does not work in `documentstyle' env

\begin{document} 				%%% revtex4

\title{
Stochastic Variational Search for \HIVLL  
}
\author{H.~{Nemura}} 		%%% revtex4
\author{Y.~{Akaishi}} 		%%% revtex4
\affiliation{ 					%%% revtex4
Institute of Particle and Nuclear Studies, KEK, Tsukuba 305-0801, Japan
} 						%%% revtex4
\author{Khin~Swe~Myint} 		%%% revtex4
\affiliation{ 					%%% revtex4
Department of Physics, Mandalay University, Mandalay, Union of Myanmar 
} 						%%% revtex4

\date{\today}

\begin{abstract} 				%%% revtex
 A four-body calculation of the $pn\Lambda\Lambda$ bound state, \HIVLL, 
 is performed using the stochastic variational method and 
 phenomenological 
 potentials. 
 The $NN$, \LN, and \LL\ potentials are taken from a recent 
 Letter by Filikhin and Gal, PRL{\bf 89}, 172502 (2002). %~\cite{FGH4LL}. 
 Although their Faddeev-Yakubovsky calculation found no 
 bound-state solution over a wide range of \LL\ interaction strengths, 
 the present variational calculation gives a bound-state energy, 
 which is clearly lower than the $\mbox{\HIIIL}+\Lambda$ threshold, 
 even for a weak \LL\ interaction strength deduced from a recent 
 experimental \BLL(\HeVILL) value. 
 The binding energies obtained are close to, and slightly larger than, 
 the values obtained from the three-body $d\Lambda\Lambda$ model in the 
 Letter. %Ref.~\cite{FGH4LL}. 
\end{abstract} 					%%% revtex

 %%% revtex4

\maketitle 					%%% ptptex \& revtex4

In a recent Physical Review Letter~\cite{FGH4LL}, 
Filikhin and Gal (FG) described systematic Faddeev-Yakubovsky (FY) 
calculations 
for the mass number $A=4$, strangeness $S=-2$ problem, in which they 
searched for 
a particle-stable bound-state of \HIVLL. 
They did not obtain a bound-state solution, 
even for a strongly attractive \LL\ interaction, 
the scattering length of which is about $a_{\Lambda\Lambda}\sim-3$ fm. 
On the other hand, they also studied the same system by using a 
three-body 
$d\Lambda\Lambda$ model, where the $\Lambda d$ interaction was 
constructed to reproduce the low-energy parameters of a 
$\Lambda pn$ Faddeev calculation for both the spin-doublet and quartet 
states. 
In contrast with the four-body $pn\Lambda\Lambda$ calculation that 
produced no bound state, the three-body $d\Lambda\Lambda$ model 
produced a particle-stable bound-state. 
One may think that this incompatibility raises an interesting problem 
concerning  
``{\it the formal relationship between these four-body and three-body 
models which do not share a common hamiltonian}.''\cite{FGH4LL} 
However, we are doubtful that there is really no bound state 
in the four-body \pnLL\ calculation. 

A recent experimental report\cite{Nagara} on the observation of \HeVILL\
in the KEK-E373 hybrid emulsion experiment 
has had 
a significant 
impact 
on strangeness nuclear physics. 
The {\it Nagara} event 
provides unambiguous identification of \HeVILL\ production, and suggests that 
the \LL\ interaction strength is rather weaker than that expected from an 
older experiment\cite{Prowse}. 

Before the publication of the Nagara event, 
we had 
already attempted 
to search for \HIVLL\ theoretically by performing a complete four-body 
calculation using a variational method~\cite{Nakaichi,Nem00}. 
The \LL\ interaction used in those studies was strongly attractive 
with a scattering length of $a_{\Lambda\Lambda}\sim -3$ fm. 
We concluded that \HIVLL\ is particle stable provided that the \LL\ 
interaction is so strong. 

The variational calculation gives an upper bound on the energy 
eigenvalue. 
In contrast with the FY formulation, although the variational 
calculation does not necessarily take the exact boundary condition into 
account, the variational principle guarantees 
that the energy obtained 
comes close to the exact value from above as the trial function is 
improved. 
Therefore, 
starting from an identical hamiltonian for the four-body %\pnLL\ 
system, 
if the bound-state solution is obtained in a variational calculation,  
the exact eigenenergy must be lower than that and the FY 
calculation should achieve this kind of solution. 

In the calculation of this four-body system, determining the \LN\ 
interaction is very important. 
Particularly, the strength in the $^3S_1$ channel of the \LN\ interaction is 
crucial, 
as well as the strength of the \LL\ interaction, in determining whether 
\HIVLL\ is particle-stable. 

The purpose of this article is twofold: 
One is to examine the recent result of the four-body  
calculation for $pn\Lambda\Lambda$ by FG. 
Our four-body calculation gives quite a different result from that of FG, 
and we discuss the structural aspects of \HIVLL\ as a four-body 
system. 
Another purpose is to clarify 
the importance 
of the 
choice of the \LN\ potential in searching for \HIVLL.

In Ref.~\cite{FGH4LL}, 
FG used phenomenological $NN$, \LN, and \LL\ potentials, which have 
functional forms of a three-range Gaussian. 
The $NN$ potential utilized in the $pn$ spin-triplet channel is 
consistent with the \HII\ binding energy, and the 
\LN\ potential is parametrized by fitting the low-energy scattering 
parameters for the Nijmegen soft-core 97f (or 97e) potential. 
For the \LL\ interaction, since there is no direct information from 
experiments in free space, FG used various parameter sets. 
A promising one, deduced by reproducing the experimental binding 
energy of \HeVILL~\cite{Nagara} 
from an $\alpha+2\Lambda$ three-body model, 
is weakly attractive 
with a scattering length of $a_{\Lambda\Lambda}=-0.77$ fm. 
For all of these interactions, 
the strength and range parameters were determined so as to be 
appropriate for $S$-wave interactions. 
We thus assume that these interactions are valid only for the 
even-partial wave component of the baryon-baryon interaction in the three- 
and four-body systems. 

For systematic calculations of \HII, \HIIIL, \HIIILs, and \HIVLL, 
we use the same sets of $NN$, \LN, and \LL\ interactions as FG used. 
The Set A \LN\ potential from Ref.~\cite{Nem00}, which has 
a different strength in the $^3S_1$ channel, is also used. 
The parameters of the Set A \LN\ potential were determined 
phenomenologically in order to reproduce the $A=3,4$ single-$\Lambda$ 
hypernuclei. %\cite{Nem00}. 
\begin{figure}[]
 \includegraphics[height=.35\textwidth]{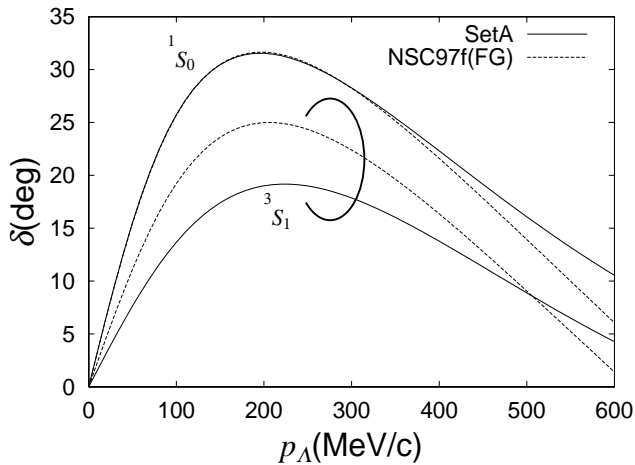} %{Fig/phsft_tgif}
 \caption{$^1S_0$ and $^3S_1$ phase shifts of \LN\ scattering as 
 a function of the $\Lambda$ momentum, $p_\Lambda$. 
 The solid lines are obtained from the Set A potential in 
 Ref.~\cite{Nem00}, the dashed lines from NSC97f(FG) in 
 Ref.~\cite{FGH4LL}. 
 }
 \label{phsftLN}
\end{figure}
Figure~\ref{phsftLN} shows the \LN\ $S$-wave phase shifts. 
In the low energy region, the $^1S_0$ phase shifts obtained from NSC97f(FG) 
and from Set A are almost identical. 
On the other hand, the $^3S_1$ interaction of the NSC97f(FG) is more 
attractive than that of Set A. 
As we show later, both \LN\ potentials reproduce the experimental 
\BL(\HIIIL) value, 
because \BL(\HIIIL) is sensitive to the strength of the $^1S_0$ \LN\ 
interaction, while it is insensitive to the $^3S_1$ strength of the \LN\ 
interaction.  
In other words, the experimental information for $A=3$ cannot determine 
the $^3S_1$ strength of the \LN\ interaction. 
Therefore, the \LN\ interaction used in the calculation of \HIVLL\ has 
to be tested not only for \BL(\HIIIL), but also for another \BL\ 
which is sensitive to the strength of the $^3S_1$ \LN\ interaction; 
for example, one can use \BL(\HIVL) and \BL(\HIVLs). 
This is one of the most important points in this article, because the 
calculated \BLL\ value is very sensitive to the choice of the \LN\ 
interaction, particularly the strength in the $^3S_1$ channel. 

In order to check the validity of the choice of the \LN\ potential, 
we calculate $A=3,4$ $\Lambda$ hypernuclei, using the NSC97f(FG) or 
the Set A \LN\ potential. 
Only for this task, the Minnesota potential\cite{Minnesota} is used for the 
$NN$ interaction. 
The parameters of the Minnesota potential were determined so as to 
reproduce low-energy $NN$ scattering data. 
The Minnesota potential reproduces reasonably well both the binding 
energies and sizes of few-nucleon systems, such as \HII, \HIII, \HeIII, 
and \HeIV\cite{VS95}.

In this work, the few-body calculations of the various systems are 
performed using 
the stochastic variational method (SVM) with correlated Gaussian (CG) 
basis functions~\cite{SVM}. 
The trial function is given by a combination of basis functions: 
\begin{equation}
\Psi_{JMTM_T} = \sum_{k=1}^{K} c_k 
{\cal A}\left[G({\bfi x}, A_k)\chi_{kJM}\eta_{TM_T}\right]. 
\label{DEFOFWF}
\end{equation}
Here, ${\cal A}$ is an antisymmetrizer acting on identical baryons, 
${\bfi x}\!=\!({\bfi x}_1, \cdots, {\bfi x}_{A-1})$ stands for a set of 
relative coordinates, 
and $\chi_{kJM}$  $\left(\eta_{TM_T}\right)$ is the spin (isospin) 
function. 
The spatial part of the trial function, $G({\bfi x}, A)$, is the 
CG, which is defined by 
\begin{eqnarray}
G({\bfi x}, A_k) &=& \exp\Big\{-\frac{1}{2}
 \sum_{i<j}^A\alpha_{kij}({\bfi r}_i-{\bfi r}_j)^2\Big\}\\ 
 &=& \exp\Big\{-{1\over 2}
  \sum_{i,j=1}^{A-1}{(A_k)}_{ij}\,{\bfi x}_i\cdot{\bfi x}_j\Big\}. 
\end{eqnarray}
The $(A-1)\times(A-1)$ symmetric matrix $A_k$ contains $A(A-1)/2$ 
independent matrix elements, which characterizes the CG basis and is 
uniquely determined in terms of $\alpha_{kij}$. 
A set of linear variational parameters $(c_1,\cdots,c_{K})$ 
is determined by using the Ritz variational principle. 
The variational parameters are optimized by 
a stochastic procedure. 
This is entirely the same as in a previous study~\cite{Nem00}. 
The reader is referred to Refs.~\cite{SVM,Nem00} for details of the 
calculation.

Before showing the results of our four-body calculations for \HIVLL, 
we report results for the binding energies of \HII, \HIIIL, and \HIIILs\ 
using the same potentials as were used by FG.
Using the triplet $pn$(FG) and NSC97f(FG) \LN\ potentials, the calculated 
binding energies were $B(\mbox{\HII})=2.250$ MeV,
$B_\Lambda(\mbox{\HIIIL})=0.237$ MeV, 
and $B_\Lambda(\mbox{\HIIILs})=0.010$ MeV. 
These energies for the three-body systems are consistent with 
those quoted by FG, although each energy is actually 
slightly larger than that of FG. 
We think that these small discrepancies are due to the $s$-wave 
approximation of the Faddeev calculation. 
Note that both calculations for \HIIILs\ produce a weakly bound state;  
this means that the SVM with CG 
basis functions and the Faddeev calculation with the $s$-wave approximation 
do work well even for the very weakly bound-state problem. 

\begin{figure}[]
 \includegraphics[height=.35\textwidth]{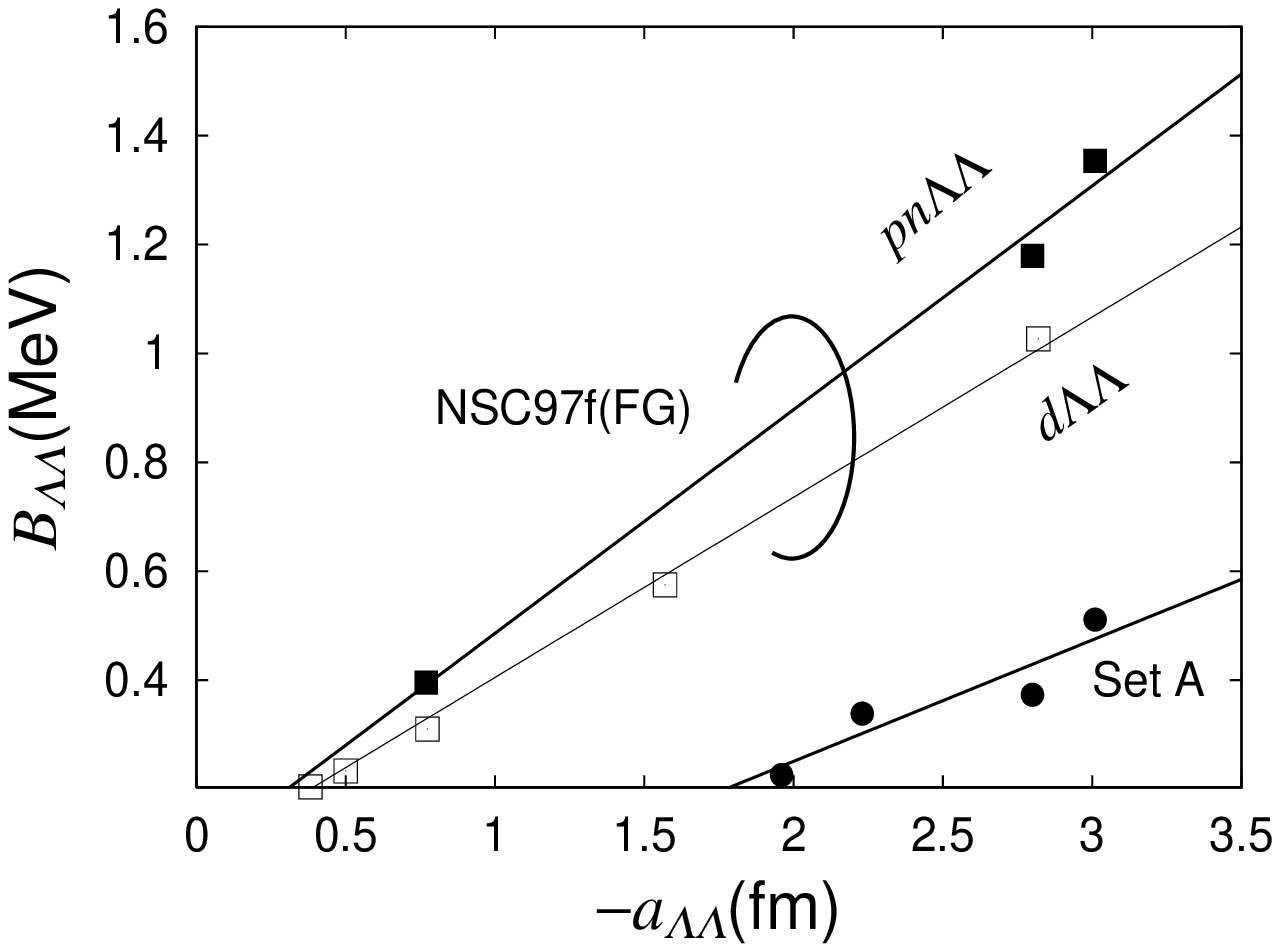} %{Fig/BLL_vs_aLL_tgif}
 \caption{Calculated \BLL(\HIVLL) as a function of the scattering 
 length, $a_{\Lambda\Lambda}$.  
 The solid squares were obtained using the NSC97f(FG) \LN\ potential and 
 the solid circles by the Set A potential. 
 The open squares are the result of the $d\Lambda\Lambda$ three-body model, 
 taken from Ref.~\cite{FGH4LL}.  
 The straight lines were drawn only for the sake of a guide to the reader.  
 }
 \label{BLLvsaLL}
\end{figure}

According to our previous studies~\cite{Nakaichi,Nem00}, 
\HIVLL\ should have a particle-stable bound state with an isospin of 
$I=0$, and an angular momentum and parity such that $J^\pi=1^+$, 
provided that a strongly attractive \LL\ interaction 
with a scattering length of $a_{\Lambda\Lambda}\sim-3$ fm, is used.  
Using such a strong \LL\ interaction, we have obtained a bound-state 
solution for \HIVLL\ (see Fig.~\ref{BLLvsaLL}). 
The \BLL(\HIVLL) values ($\gsim 1.2$ MeV) obtained are more than two-times 
larger than the values obtained in our previous studies ($\sim 0.5$ MeV) with the 
Set A \LN\ potential. 
This is due to the difference in the strength of the \LN\ interaction 
in the $^3S_1$ channel (see Fig.~\ref{phsftLN}). 

\begin{figure}[]
 \includegraphics[height=.35\textwidth]{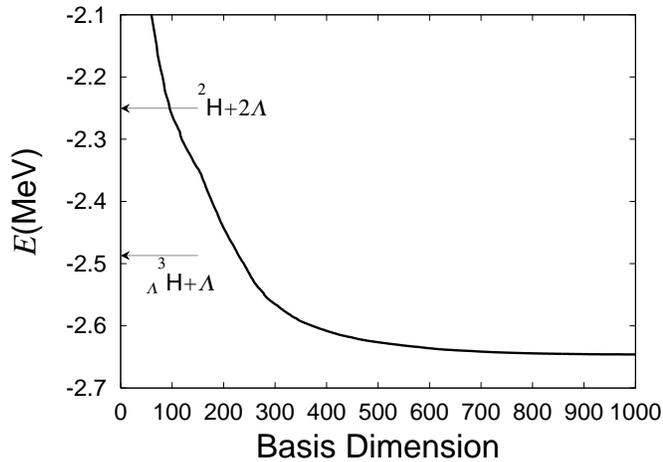} %{Fig/ene_H4LL_3}
 \caption{Energy convergence of \HIVLL\ as a function of the basis 
 dimension, $K$. 
 The interactions are taken from Ref.~\cite{FGH4LL}, 
 spin-triplet $pn$, NSC97f(FG) \LN, and \LL\ deduced from the recent 
 experimental \BLL(\HeVILL). 
 The converged energy is clearly lower than the $\mbox{\HIIIL}+\Lambda$ 
 threshold. 
 }
 \label{ENEofH4LL}
\end{figure}

The four-body calculation using a weaker \LL\ interaction 
($a_{\Lambda\Lambda}=-0.77$ fm) is a 
challenging problem, since the three-body $d\Lambda\Lambda$ model 
by FG predicts a particle-stable bound-state with a very small binding 
energy ($\mbox{\BLL}\sim 0.3$ MeV). 
For such a weakly bound four-body calculation, though the convergence 
of the energy is rather slow, the energy obtained is clearly lower than the 
$\mbox{\HIIIL}+\Lambda$ threshold, and we found that the ground 
state is particle stable (see Fig.~\ref{ENEofH4LL}). 
This is a genuine four-body calculation, and 
the calculated $\mbox{\BLL(\HIVLL)}\sim 0.4$ MeV is slightly larger 
than that from the $d\Lambda\Lambda$ three-body calculation by FG, as 
shown in Fig.~\ref{BLLvsaLL}. 

As can be seen in Fig.~\ref{BLLvsaLL}, the difference in the \BLL\ values 
between the present four-body model and the FG 
three-body model becomes larger as the strength of the \LL\ interaction 
increases. 
Moreover, the two lines (labelled ``$pn\Lambda\Lambda$'' and 
``$d\Lambda\Lambda$'') in Fig.~\ref{BLLvsaLL} seem to meet each other at 
the point where $a_{\Lambda\Lambda}=0$ fm. 
This means that the polarization of the $pn$ subsystem is small, and that the 
$d\Lambda\Lambda$ model is a good approximation if the \LL\ interaction 
is very weak. 
The polarization of the deuteron subsystem grows as 
the strength of the \LL\ interaction increases. 
\begin{table}[]
 \begin{tabular}{lccccc}
  \hline 
  & $\langle T_c\rangle$ & $\langle V_{NN}\rangle$ & $E_c$ &
  $\sqrt{\langle r_{NN}^2\rangle}$ & $\sqrt{\langle r_{\Lambda N}^2\rangle}$ \\ 
  \hline
  \ \ \HII & 18.74 & $-20.99$ & $-2.25$ & 3.85 \\
  \ \ \HIIILs & 19.09 & $-21.20$ & $-2.12$ & 3.79 & 37.8 \\
  \ \ \HIIIL &  20.70 & $-22.30$ & $-1.59$ & 3.54 & 8.88 \\
  \HIVLL ($a_{\Lambda\Lambda}\!=\!-0.77$fm) & 22.28 & $-23.17$ & $-0.88$ & 3.34 & 7.92 \\
  \HIVLL ($a_{\Lambda\Lambda}\!=\!-2.8$fm) &  24.73 & $-24.55$ & $\ \ 0.18$ & 3.08 & 4.82 \\
  \hline 
 \end{tabular}
 \caption{Energy expectation values of kinetic ($T_c$) and  
 potential ($V_{NN}$) terms, and the sum of these energies ($E_c$), 
 for the $pn$ subsystem, 
 in units of MeV. 
 The rms distance between a proton and a neutron, or between a nucleon and 
 a $\Lambda$, is also listed, in units of fm. 
 The spin-triplet $pn$ and NSC97f(FG) \LN\ potentials, taken from 
 Ref.~\cite{FGH4LL},  were used. 
 }
 \label{NNrearr}
\end{table}
Table~\ref{NNrearr} lists the energy expectation values for the proton 
and neutron subsystem in each (hyper) nucleus, and also the 
root-mean-square distances between a $p$ and an $n$, or between a nucleon and 
a $\Lambda$. 
Here, $T_c$ is the kinetic energy of the $pn$ subsystem, which is defined by 
$T_c={({\bfi p}_1-{\bfi p}_2)^2/4m_N}.$
The table shows that the influence of the $\Lambda$ particle upon the 
internal structure of the $pn$ subsystem becomes large as the $\Lambda$  
particle comes close to the nucleon. 
Especially in the case of a strongly attractive \LL\ potential, 
the change in the internal energy ($E_c$) or of the rms distance 
($\sqrt{\langle r_{NN}^2\rangle}$) is significant.

As can be seen in Fig.~\ref{BLLvsaLL}, the \BLL\ value is sensitive to the 
choice of the \LN\ potential. 
For the purpose of predicting whether \HIVLL\ exists as a 
particle-stable bound-state, the \LN\ potential has to be examined 
carefully. 
\begin{table}[]
 \begin{tabular}{llll}
  \hline 
  & \BL(\HIIIL) & \BL(\HIVL) & \BL(\HIVLs) \\ 
  \hline 
  NSC97f(FG) & 0.24 & 2.69 & 1.99 \\ 
  Set A      & 0.18 & 2.24 & 1.14 \\ 
  Experiment & $0.13\pm0.05$ & $2.04\pm0.04$ & $1.00\pm0.04$ \\
  \hline 
 \end{tabular}
 \caption{$\Lambda$ separation energies, given in units of MeV, 
 of $A=3,4$ single-$\Lambda$ hypernuclei. 
 The Minnesota $NN$ potential was used. 
 }
 \label{BL}
\end{table}
Table~\ref{BL} compares the \BL\ values of $A=3,4$ hypernuclei. 
The calculated \BL\ value of the $A=4$ system using NSC97f(FG) is larger 
than that using Set A, or larger than the experimental value. 
Particularly, the $^3S_1$ strength of NSC97f(FG) is apparently too 
strong to reproduce the \BL(\HIVLs) value, though the NSC97f(FG) 
reproduces reasonably well the \BL(\HIIIL) value. 
It would, therefore, be rash to conclude that \HIVLL\ has a particle-stable 
bound-state, though the present four-body calculation with the 
NSC97f(FG) gives a bound-state solution, even for a weaker \LL\ 
interaction, such as $a_{\Lambda\Lambda}\!\!=\!\!-0.77$ fm.

The present four-body calculation gives quite a different result from that 
of the FY study discussed in Ref.~\cite{FGH4LL}. 
At present, we have no clear explanation for why the FY search for \HIVLL\ has 
not found a bound-state solution. 
We also checked the accuracy of the present variational calculation by 
examining the virial theorem\cite{SVM}; 
For an exact eigenstate of the hamiltonian, $H=T+V$, we have 
\begin{equation}
\langle T\rangle={1\over 2}\langle W\rangle, \quad \mbox{with}\quad 
W=\sum_{i=1}^A{\bfi r}_i\cdot{\partial V\over \partial {\bfi r}_i}. 
\end{equation}
For the four-body calculation, we obtained the ratio 
$2\langle T\rangle/\langle W\rangle=1.000016$ for a 
weak \LL\ potential ($a_{\Lambda\Lambda}=-0.77$ fm), and the ratio 
$0.999962$ for a strong \LL\ potential ($a_{\Lambda\Lambda}=-2.8$ fm).  
Therefore, we think that the present four-body calculation gives a 
virtually exact eigenenergy, and that 
the \BLL\ value obtained by a four-body calculation for $pn\Lambda\Lambda$  
should be close to (and slightly larger than) the energy given by the 
$d\Lambda\Lambda$ three-body model.  
In comparison with the $d\Lambda\Lambda$ three-body model (in 
Fig.~\ref{BLLvsaLL}), 
the present 
result seems to be reasonable, in contrast to that 
of the FY four-body calculation in Ref.~\cite{FGH4LL}. 

The contribution to the binding energy from the higher partial wave 
components are marginal. 
The present potentials are all central and have Gaussian form factors. 
This Gaussian radial form 
(e.g., $Ve^{-\kappa ({\bfi r}_i-{\bfi r}_j)^2}$) is rewritten so as to 
be valid for each angular momentum in terms of nonlocal 
potentials:\cite{LLXN} 
\begin{equation}
 \begin{array}{rl}
Ve^{-\kappa ({\bfi r}_i-{\bfi r}_j)^2}& \!\!=\int d{\bfi r}d{\bfi r}^\prime 
|\delta({\bfi r}_i-{\bfi r}_j-{\bfi r}^\prime)\rangle
\langle\delta({\bfi r}_i-{\bfi r}_j-{\bfi r})| \\
&\times
\sum_{l=0}^\infty Ve^{-\kappa r^2} {\delta(r^\prime-r)\over r^2}
\sum_{m} Y_{lm}^\ast(\hat{\bfi r}^\prime)Y_{lm}(\hat{\bfi r}).
 \end{array}
\end{equation}
We also calculated the binding energies in which the $NN$, \LN, and \LL\ 
potentials are 
restricted to be valid only for the $l=0$ component. 
The binding energy calculated is $B(\mbox{\HIVLL})=2.382$ MeV 
($a_{\Lambda\Lambda}=-0.77$ fm), or $B(\mbox{\HIVLL})=2.842$ MeV 
($a_{\Lambda\Lambda}=-2.8$ fm). 
Each energy is still lower than the $\mbox{\HIIIL}+\Lambda$ 
threshold (For $l=0$ truncated interactions, we obtained 
$B(\mbox{\HIIIL})=2.365$ MeV). 

We should emphasize that in the study of \HIVLL; 
the $^3S_1$ \LN\ interaction has to be determined very carefully, since 
\BLL\ is sensitive to the $^3S_1$ channel of the \LN\ interaction.  
Therefore, a check of the \LN\ potential concerning the observed binding energy of 
only the subsystem, \HIIIL\ $({1\over 2}^+)$, is 
insufficient. 

One might think that the spin-doublet structure of $A=4$ 
$\Lambda$ 
hypernuclei is a means of determining the $^3S_1$ 
\LN\ interaction. 
However, this strategy without any explicit $\Sigma$ admixture would lead us 
to a serious problem concerning the $A=5$ anomaly~\cite{DHT,Nem00,Survey}. 
According to recent studies, taking account of the explicit $\Sigma$ degrees 
of freedom~\cite{Aka00,abinitio,Miyagawa,Hiyama,Nogga}, the \LNSN\ 
coupling, plays a crucial role in the binding mechanism of $s$-shell 
$\Lambda$ hypernuclei.   
In other words, even the spin-doublet structure of $A=4$ 
$\Lambda$ 
hypernuclei does not pin down the $^3S_1$ \LN\ interaction, 
and the \LN\ potential used in the study of \HIVLL, has to be tested on 
a complete set of the observed $s$-shell $\Lambda$ hypernuclei. 
Moreover, the $\alpha+2\Lambda$ three-body model might be inappropriate for 
deducing the \LL\ interaction in free space from the recent 
experimental 
information on \BLL(\HeVILL), since \LNSN\ coupling plays an 
important role even for \HeVL, and the rearrangement energy of the core 
nucleus (\HeIV) is rather large~\cite{abinitio,rearrange}. 
Therefore, a study aimed at searching for \HIVLL\ needs not only a four-body 
calculation, but also five-body (\HeVL) and six-body (\HeVILL) 
calculations. 
Furthermore, $\Lambda\Lambda-\Xi N$ coupling effects should be 
explicitly taken into account, because the Pauli suppression effect in 
the $\Xi$ channel of \HeVILL\ is appreciably large\cite{LLXN}. 
A theoretical search for \HIVLL\ is still an open subject.

\begin{acknowledgments}         %%% revtex
The authors are thankful to 
T. Harada for useful communications.  
One of the authors (H.N.) would like to thank for 
JSPS Research Fellowships for Young Scientists. 
\end{acknowledgments}          %%% revtex

\end{document}